\documentclass[12pt]{article}
\usepackage{epsfig}
\usepackage{amsmath}
\usepackage{amsfonts}

\begin{document}
\pagestyle{empty}
\begin{flushright}
UMN-TH-2608/07
\end{flushright}
\vspace*{5mm}

\begin{center}
{\Large\bf Holographic Mixing Quantified}
\vspace{1.0cm}

{\sc Brian Batell}\footnote{E-mail:  batell@physics.umn.edu}
{\small and}
{\sc Tony Gherghetta}\footnote{E-mail:  tgher@physics.umn.edu}
\\
\vspace{.5cm}
{\it\small {School of Physics and Astronomy\\
University of Minnesota\\
Minneapolis, MN 55455, USA}}\\
\end{center}

\vspace{1cm}
\begin{abstract}
We compute the precise elementary/composite field content of mass eigenstates in holographic
duals of warped models in a slice of AdS$_5$. This is accomplished by decomposing the bulk
fields not in the usual Kaluza-Klein basis, but rather into  a holographic basis of 4D fields,
corresponding to purely elementary source or CFT composite fields.
Generically, this decomposition yields kinetic and mass mixing between the elementary and composite sectors of the holographic theory.
Depending on where the bulk zero mode is localized, the elementary/composite content may differ radically, which we show explicitly for several examples including the bulk Randall-Sundrum graviton, bulk gauge boson, and Higgs boson.
\end{abstract}

\vfill
\begin{flushleft}
\end{flushleft}
\eject
\pagestyle{empty}
\setcounter{page}{1}
\setcounter{footnote}{0}
\pagestyle{plain}

\section{Introduction}

The AdS/CFT correspondence suggests that certain strongly coupled four-dimensional (4D) gauge theories are dual to weakly coupled theories defined on a five-dimensional (5D) warped geometry \cite{ads1,ads2,ads3}. The extra dimension can therefore be utilized as a calculational tool to understand properties of composite states in the 4D theory.
Indeed the dual interpretation of the compact Randall-Sundrum model \cite{rs} is that the Standard Model fields localized on the infrared (IR) brane are composite states \cite{pheno1,pheno2,pheno3}.

In a slice of 5D anti-de Sitter (AdS) space, the finite ultraviolet (UV) boundary in the warped extra dimension translates into a UV cutoff of the dual conformal field theory (CFT), and in turn implies the existence of a dynamical source field. This is an elementary degree of freedom, external to the CFT. The presence of an IR brane is interpreted as a spontaneous breakdown of conformal symmetry, marked by the appearance of resonances, or bound states of fundamental CFT fields. Mixing between the elementary (source) and composite (CFT) sectors produces the mass eigenstates of the theory, corresponding to the Kaluza-Klein fields from the 5D perspective. In other words, the mass eigenstates in the dual theory exhibit partial compositeness.
A complete, quantitative treatment of this mixing has thus far not been presented. For instance, it has not been possible to say precisely how much a given mass eigenstate is composed of source and CFT fields. 

The purpose of this paper is to quantitatively characterize the mixing between the elementary and composite sectors in holographic duals of theories defined on a slice of AdS$_5$. Instead of compactifying the 5D theory using a Kaluza-Klein decomposition, which results in a diagonal action, we propose to expand the bulk field directly in terms of purely source and CFT bound states. We designate the term {\it holographic basis} to denote the set of fields and $y$-dependent profiles that form this expansion. The decomposition generically results in both kinetic and mass mixing. We set up the general eigenvalue problem and outline how to diagonalize the system. Of course, this must lead back to the Kaluza-Klein, or mass eigenbasis, which we show explicitly for numerous examples. The transformation that diagonalizes the system tells us precisely how much a particular field is composed of elementary and composite degrees of freedom. Geometrical intuition of phenomenological models in warped space can now clearly be translated into the language of elementary/composite mixing. This formalism is applicable to any bulk theory which contains a massless mode in the Kaluza-Klein decomposition.

The theory we study in this paper is a 5D scalar field with bulk and boundary masses \cite{gp,review}. By tuning these masses, a zero mode can be localized arbitrarily in the extra dimension, corresponding to different dual interpretations of the bulk theory. For particular values of the boundary mass, this theory mimics that of a bulk graviton or gauge boson, and is thus a simple but relevant example to study. This formalism can also be applied to the case of bulk fermions \cite{inprep}. A phenomenological approach to holographic mixing has recently been applied to warped phenomenology in Ref.~\cite{crss}.

The structure of this paper is as follows: In Section 2 we begin with a truncated $2\times 2$ mixing problem which illustrates simply many of the nonstandard features of this eigenvalue problem, including kinetic mixing and nonorthogonal transformations. We review the theory of a 5D scalar field with bulk and boundary masses in Section 3 in particular focusing on those aspects of the dual interpretation which will provide the motivation for the holographic basis. In Section 4, we propose the holographic basis and analyze the general eigenvalue problem. In Section 5 the specific examples of the graviton and gauge boson, as well as generic composite scalar fields, are presented, showing explicitly the elementary/composite content of each field. Conclusions and possible directions for future work are presented in Section 6.

\section{A truncated holographic mixing problem}

Many features of the holographic eigenvalue problem are unfamiliar. As mentioned in the introduction, there is always kinetic mixing between the source and CFT sectors. Diagonalizing the system thus requires the intermediate step of canonical normalization, which is simply a rescaling of the fields. As a result, the transformation matrix that diagonalizes the system is not orthogonal. It is still straightforward to specify the elementary/composite content of a particular mass eigenstate by examining the corresponding eigenvector. Before proceeding to the general problem in which we analyze the entire tower of composite states, it is instructive to consider a truncated version which illustrates the unfamiliar aspects of the problem. Of course, these issues have been encountered before. For instance, kinetic mixing between the photon and $Z$ boson occurs through electroweak corrections. Another example is given by $Z-Z'$ boson mixing, where in general there may exist mass as well as kinetic mixing \cite{zzp}.

Forseeing our main results, let us consider a truncated holographic theory containing a massless source field $\varphi^s(x)$ and a single composite field $\varphi^1(x)$ with mass $M_1$. Kinetic mixing in the Lagrangian implies that the mass eigenstates are elementary/composite mixtures. The Lagrangian is
\begin{equation}
{\cal L}={\cal L}_{elementary}+{\cal L}_{composite}+{\cal L}_{mix},
\end{equation}
where
\begin{eqnarray}
{\cal L}_{elementary}&=& -\frac{1}{2}(\partial_\mu \varphi^s)^2, \\
{\cal L}_{composite}&=& -\frac{1}{2}(\partial_\mu \varphi^1)^2-\frac{1}{2}M_1^2( \varphi^1)^2,\\
{\cal L}_{mix}&=&-\sin\theta~ \partial_\mu\varphi^s \partial^\mu \varphi^1.
\end{eqnarray}
The kinetic mixing is parameterized by the mixing angle $\theta$. Notice that we neglect any mass mixing in ${\cal L}_{mix}$. This is in fact a realistic assumption; as we will see later, there is no mass mixing for many phenomenological examples. Actually, this truncated example quite accurately describes the bulk Randall-Sundrum graviton and gauge boson, discussed in Section 5.  

We diagonalize the system in three steps. First, we perform an orthogonal rotation which leaves the kinetic terms diagonal, but induces a mass mixing between the rotated fields. Next, the kinetic terms are  canonically normalized by scaling the fields. Finally, we rotate the scaled fields to diagonalize the mass terms. Altogether, the system is diagonalized by the following field redefinition:
\begin{equation}
\left( \begin{array}{c} \varphi^s  \\ \varphi^1 \end{array} \right)\rightarrow
\left( \begin{array}{c} \phi^0  \\  \phi^1 \end{array} \right)
=
\left( \begin{array}{cc} 1 & \sin \theta   \\ 0 & \cos \theta \end{array} \right)
 \left( \begin{array}{c} \varphi^s   \\ \varphi^1 \end{array} \right). \\
\label{tran2by2}
\end{equation}
The physical Lagrangian is thus
\begin{equation}
{\cal L} = -\frac{1}{2}(\partial_\mu \phi^0)^2 -\frac{1}{2}(\partial_\mu \phi^1)^2 -\frac{1}{2} M_1^2 \sec^2 \theta (\phi^1)^2.
\end{equation}
There is a massless eigenstate $\phi^0(x)$ as well as a massive state $\phi^1(x)$, corresponding to Kaluza-Klein modes in the 5D warped theory.

The transformation (\ref{tran2by2}) is not orthogonal. This is easy to understand: the intermediate step of canonical normalization can be seen as a redefinition of the fields via a diagonal nonorthogonal matrix. Still, the elementary/composite content of each mass eigenstate can easily be read off from (\ref{tran2by2}). For example, the fraction $\epsilon$ of the zero mode $\phi^0(x)$ that is composite is
\begin{equation}
\epsilon=
\frac{\sin^2\theta}{1+\sin^2\theta}~.
\end{equation}
Another interesting feature is that only the massless mode $\phi^0(x)$ contains the elementary source field; the massive mode is purely composite. We will see that this feature occurs for a wide class of examples in the general problem.

At this point, we might ask why the mixing in the original theory needs to be formulated in terms of kinetic mixing instead of the more standard mass mixing. After all, we can always transform to a set of fields where only mass mixing occurs.  Indeed, this is exactly what the first two steps in our diagonalization procedure accomplishes. However, there is a physical reason we must work with kinetic mixing: the holographic interpretation dictates that the pure source and CFT states have a particular set of diagonal mass terms. We will show that the correct diagonal masses occur in a basis where there is kinetic mixing. In fact, kinetic mixing was anticipated in the case of the bulk gauge field in \cite{pheno1}.

\section{The Kaluza-Klein mass eigenbasis}

Let us now turn to theories in a slice of AdS$_5$, reviewing aspects of both the conventional Kaluza-Klein analysis as well as the holographic interpretation. We will study a scalar field with bulk and boundary masses, showing how to localize a zero mode anywhere in the bulk \cite{gp, review}. The metric for this background is
\begin{equation}
ds^2=e^{-2 k y}\eta_{\mu\nu} dx^\mu dx^\nu+dy^2~,
\end{equation}
where $k$ is the AdS curvature scale. The extra coordinate ranges from $y=0$ to $y=\pi R$ where there exists a UV and IR brane, respectively.
Latin letters ($A,B,\dots$) denote 5D indices, while Greek letters ($\mu,\nu,\dots$) are reserved for 4D indices. 4D indices are raised and lowered with $\eta={\rm diag}(-,~+,~+,~+)$.

Consider the action describing a real scalar field $\phi(x,y)$ propagating on this background:
\begin{equation}
S=\int d^5x \sqrt{-g}\left[ -\frac{1}{2}(\partial_M \phi)^2-\frac{1}{2}ak^2\phi^2-bk\phi^2\left(\delta(y)-\delta(y-\pi R)\right)\right],
\label{a1}
\end{equation}
where the bulk and boundary masses are written in terms of the AdS curvature scale $k$ with dimensionless parameters $a$ and $b$, and $\phi(x,y)$ satisfies the boundary condition
\begin{equation}
(\partial_5-bk)\phi(x,y)\bigg\vert_{0,\pi R}=0~.
\label{bc}
\end{equation}
The standard procedure to obtain the 4D modes is to perform a
Kaluza-Klein decomposition,
\begin{equation}
\phi(x,y)=\sum_{n=0}^\infty \phi^n(x)f^n(y),
\label{kk}
\end{equation}
where the resulting 4D theory is diagonal in the Kaluza-Klein states, and is thus written in a mass eigenbasis.
The eigenfunctions $f^n(y)$ are orthonormal,
\begin{equation}
\int_0^{\pi R} dy ~e^{-2 k y}f^n f^m=\delta^{n m}~,
\label{norm}
\end{equation}
and satisfy the equation of motion,
\begin{equation}
\Big[\partial_5e^{-4 k y}\partial_5-ak^2e^{-4 k y}\Big]f^n(y)=-m_n^2e^{-2 k y}f^n(y)~,
\label{eom1}
\end{equation}
with the boundary conditions (\ref{bc}).

In order for a massless  zero mode to exist $(m_0=0)$ the mass parameters are tuned to satisfy the relation
\begin{equation}
b=2\pm\alpha\equiv 2\pm\sqrt{4+a}~,
\label{tune}
\end{equation}
where $\alpha$ is taken to be real, implying that $a\geq-4$ and $ -\infty<b<\infty$.
The normalized zero mode solution compatible with the boundary condition (\ref{bc}) is then given by
\begin{equation}
f^0(y)=\sqrt{\frac{2(b-1)k}{e^{2(b-1)\pi k R}-1}}e^{b k y}~.
\label{0mode}
\end{equation}
Since $b$ can take any real value, the massless mode can be localized anywhere in the fifth dimension,
admitting qualitatively different dual interpretations of the 5D theory. In fact, as we will see later, this simple scalar field theory quantitatively captures the elementary/composite mixing of other bulk bosonic fields, in particular, the graviton ($b=0$), and the gauge boson ($b=1$),
making it a very useful and general theory to study.

We can also derive the eigenfunctions of (\ref{eom1}) for excited modes, $m_n\neq0$. The spectrum can be found by applying boundary conditions (\ref{bc}) to the eigenfunctions and is determined by the zeros of the following equation:
\begin{equation}
J_{\alpha\pm 1}\left(\frac{m_n}{k}\right)Y_{\alpha\pm 1}\left(\frac{m_n e^{\pi k R}}{k}\right)-Y_{\alpha\pm 1}\left(\frac{m_n}{k}\right)J_{\alpha\pm 1}\left(\frac{m_n e^{\pi k R}}{k}\right)=0~.
\label{spectrum1}
\end{equation}
As mentioned above, the resulting 4D action is diagonal (by construction), and thus the $\phi^n(x)$ are the physical fields in the theory.

\subsection{The holographic  dual interpretation}
Alternatively we can analyze the theory (\ref{a1}) by using the so-called holographic procedure
\cite{review}. Inspired by the AdS/CFT correspondence, the bulk field $\phi(x,y)$ has a corresponding CFT operator ${\cal O}$ in the 4D dual theory, and the UV boundary value $\varphi_0(x)=\phi(x,0)$ is a source for this operator in the partition function. The dual particles are bound states composed of fundamental fields in the CFT. Thus, we expect these states to appear as poles in the two-point function $\langle {\cal O O}\rangle$, much like mesons in QCD. Indeed, in the pure AdS case, where the dual theory is described by a gauge theory with a large number of colors $N_c$, it is well known that the correlator can be written as a sum over an infinite number of narrow resonances \cite{thooft,witten}:
\begin{equation}
\langle {\cal O O}\rangle=\sum_n\frac{F_n^2}{p^2+M_n^2}~,
\label{largeN}
\end{equation}
where $F_n=\langle 0 |{\cal O} |n\rangle$ is the amplitude for the operator ${\cal O}$ to excite a resonance from the vacuum. We use $M_n$ to denote the masses of the CFT resonances, distinct from the eigenmasses $m_n$.
The composite states are composed of the fundamental fields in the large $N_c$ gauge theory. However, with a UV cutoff the source field becomes dynamical, mixing with composite states and modifying the spectrum. The low-energy Lagrangian is not diagonal but contains mixing between the source and CFT states.

The poles corresponding to the CFT masses $M_n$ are determined from the correlator $\langle {\cal O O}\rangle$. Following the holographic procedure, the correlator is obtained by integrating out the bulk degrees of freedom, and deriving the boundary effective action.
For the action (\ref{a1}), the self-energy $\Sigma(p)$, which contains the correlator $\langle {\cal O O}\rangle$ and induces dynamics for the source, is given by~\cite{review}
\begin{equation}
\Sigma(p)=\mp i p ~\frac{
J_{\alpha\pm 1}\left(\frac{i p}{k}\right) Y_{\alpha\pm 1}\left(\frac{i p e^{\pi k R}}{k}\right)
-Y_{\alpha\pm 1}\left(\frac{i p}{k}\right) J_{\alpha\pm 1}\left(\frac{i p e^{\pi k R}}{k}\right)}
{ J_{\alpha}\left(\frac{i p}{k}\right) Y_{\alpha\pm 1}\left(\frac{i p e^{\pi k R}}{k}\right)
-Y_{\alpha}\left(\frac{i p}{k}\right) J_{\alpha\pm 1}\left(\frac{i p e^{\pi k R}}{k}\right)}~.
\label{sigma}
\end{equation}
Note that we have omitted a factor $1/g_\phi^2 k = N_c$ in our defininition of $\Sigma(p)$.
In the limit of large Euclidean momentum, we can extract the dimension $\Delta$ of the CFT operator ${\cal O}$:
\begin{equation}
\Delta=2+\big\vert b-2 \big\vert,
\label{dimO}
\end{equation}
indicating relevant, marginal, or irrelevant source/CFT mixing depending on $b$. In particular, strong mixing occurs at low energies for $1<b<3$.

The poles of $\Sigma(p)$ are given by the zeros of the denominator in (\ref{sigma}). Comparing with Eq. (\ref{largeN}), we conclude that the mass spectrum of the composite CFT states is given by
\begin{equation}
J_{\alpha}\left(\frac{M_n}{k}\right)Y_{\alpha\pm 1}\left(\frac{M_n e^{\pi k R}}{k}\right)-Y_{\alpha}\left(\frac{M_n}{k}\right)J_{\alpha\pm 1}\left(\frac{M_n e^{\pi k R}}{k}\right)=0~.
\label{spectrum2}
\end{equation}
This does not correspond to the physical spectrum of the dual theory because we have not yet taken into account the dynamics of the source field. Nontrivial mixing between the source and CFT sectors is generated through the interaction $\varphi_0 {\cal O}$ and modifies the spectrum. Quantum corrections involving insertions of the $\langle {\cal O O}\rangle$ correlator effectively invert $\Sigma(p)$. The eigenmasses $m_n$ are thus given by the zeros of $\Sigma(p)$ rather than the poles, identical to the masses of the Kaluza-Klein states (\ref{spectrum1}). Hence, the spectra of the two theories are indeed identical.

The decay constants $F_n$ are found by computing the residues of $\Sigma(p)$ at the pole $p^2=-M_n^2$ \cite{qcd1,qcd2}:
\begin{equation}
F_n=\frac{\sqrt{2 k}M_n~Y_{\alpha\pm 1}\left(\frac{M_n e^{\pi k R}}{k}\right)}{\sqrt{Y^2_{\alpha}\left(\frac{M_n}{k}\right)- Y^2_{\alpha\pm 1}\left(\frac{M_n e^{\pi k R}}{k}\right)}}~.
\label{fn1}
\end{equation}
Later, we will show that these amplitudes match those computed using the holographic basis.

There are two branches in the correlator $\Sigma(p)$: a $(+)$ branch for $b>2$, and a $(-)$ branch for $b<2$ \cite{review}. On the $(-)$ branch the dual description consists of a massless elementary source field coupled to massive composite CFT states, whereas on the $(+)$ branch, the massless particle in the dual theory is primarily a CFT bound state, while the elementary source becomes very massive. For very large $|b|$ the mixing between the elementary and CFT sectors is negligible, and the mass eigenbasis is very well approximated by the holographic basis. However near the transition point $b \sim 2$ strong mixing between the elementary and composite sectors produces the mass eigenstates.

Although exactly massless on the $(-)$ branch, the source field picks up a large mass of order $k$ on the $(+)$ branch, which can be derived by expanding $\Sigma(p)$ at high momentum $k e^{-\pi k R} < p < k$~:
\begin{equation}
M_0^2\sim4 (b-2)(b-3)k^2.
\label{M0}
\end{equation}
Furthermore, the CFT produces an exponentially light composite state below the IR scale, which can be derived from (\ref{spectrum2}) at low momentum $p < ke^{\pi k R}$,
\begin{equation}
M_1^2\sim 4 \alpha (\alpha+1)k^2 e^{-2(\alpha+1)\pi k R}= 4 (b-2) (b-1)k^2 e^{-2(b-1)\pi k R}.
\label{Mapprox}
\end{equation}
The next pole appears around the IR scale ($\sim$ TeV in RS1).
Qualitatively, this is telling us that the massless particle on the (+) branch corresponds primarily to a composite CFT bound state, and contains only a very small admixture of the elementary source field.
No such light pole appears in $\Sigma(p)$ on the ($-$) branch.  This makes sense because on the $(-)$ branch, the massless state corresponds to the elementary source field, external to the CFT, and thus should not appear as a pole in $\Sigma(p)$.
As we increase $b$ on the $(+)$ branch, the first CFT bound state becomes lighter and lighter, as can be seen from (\ref{Mapprox}), and accordingly, the first pole moves increasingly closer to $p^2=0$. This agrees with the analysis in \cite{gp03,review}, where the UV brane was removed completely and a massless pole was found.

What we will accomplish next is to write the analog of a chiral Lagrangian in QCD: an effective field theory describing a set of CFT resonances mixing with an elementary sector. The CFT bound states will have masses $M_n$ determined by (\ref{spectrum2}), the dynamical source field will be either massless or massive depending on which branch $(+/-)$ is under consideration, and there will be mixing between the source field and the CFT bound states. It is crucial to notice that the CFT spectrum (\ref{spectrum2}) would arise in the 5D theory by applying Dirichlet conditions at the UV boundary ($y=0$). This observation will play a key role in defining the holographic basis, to which we now turn.

\section{The holographic basis}

Based on the preceding discussion, it is clear that mass eigenstates in the holographic theory are a consequence of mixing between the elementary(source) and composite(CFT) sectors. 
To represent the mixing taking place between the elementary and composite sectors, we propose to decompose the action by expanding the field $\phi(x,y)$ directly in terms of a source field $\varphi^s(x)$ and a tower of CFT bound states $\varphi^n(x)$, with the associated wavefunctions $g^s(y)$ and $g^n(y)$:
\begin{equation}
\phi(x,y)=\varphi^s(x) g^s(y) +\sum_{n=1}^\infty\varphi^n(x)g^n(y)~.
\label{holobasis}
\end{equation}
We refer to this expansion as the {\it holographic basis}.

Clearly the profiles $g^s(y)$ and $g^n(y)$ must be different than the Kaluza-Klein profiles $f^n(y)$. Consider first the profiles of the CFT resonances $g^n(y)$. Recall from the previous section that the CFT spectrum (\ref{spectrum2}) derived from the correlator $\Sigma(p)$ corresponds to applying a pure Dirichlet condition at the UV boundary, $\phi(x,y=0)=0$, and the modified Neumann condition (\ref{bc}) at the IR boundary (\ref{spectrum2}). We assume therefore that the CFT profiles $g^n(y)$ satisfy the bulk equation of motion (with eigenvalues $M_n^2$) and the following boundary conditions:
\begin{eqnarray}
g^n(y)\bigg\vert_0&=&0~, \label{holobcUV}\\
(\partial_5-b k)g^n(y)\bigg\vert_{\pi R}&=&0~.
\label{holobc}
\end{eqnarray}
Also, we impose a wavefunction normalization analogous to (\ref{norm}) in order to have canonical kinetic terms.
Explicitly, the CFT eigenfunctions are given by
\begin{equation}
g^n(y)=N^{{\rm CFT}}_n e^{2 k y} \left[J_\alpha\left(\frac{M_n e^{ky}}{k}\right)+\kappa(M_n)Y_\alpha\left(\frac{M_n e^{ky}}{k}\right)\right]~.
\label{gcft}
\end{equation}
The coefficient $\kappa(M_n)$ is found by applying the boundary conditions (\ref{holobcUV}) and
(\ref{holobc}):
\begin{equation}
\kappa(M_n)=-\frac{J_{\alpha}\left(\frac{M_n}{k}\right)}
{Y_{\alpha}\left(\frac{M_n}{k}\right)}
=-\frac{J_{\alpha \pm 1}\left(\frac{M_n e^{\pi k R}}{k}\right)}
{Y_{\alpha \pm 1}\left(\frac{M_n e^{\pi k R}}{k}\right)}~,
\label{gamma}
\end{equation}
yielding the mass eigenvalue equation (\ref{spectrum2})~.
We also give here the normalization $N^{{\rm CFT}}_n$, necessary for the computation of the source/CFT mixing:
\begin{equation}
N^{{\rm CFT}}_n=\frac{\pi M_n}{\sqrt{2 k}}
\frac{ Y_{\alpha}\left(\frac{M_n}{k}\right)
Y_{\alpha\pm 1}\left(\frac{M_n e^{\pi k R}}{k}\right)}
{\sqrt{Y^2_{\alpha}\left(\frac{M_n}{k}\right)-Y^2_{\alpha\pm 1}\left(\frac{M_n e^{\pi k R}}{k}\right)}}~.
\end{equation}

Next, for the source profile $g^s(y)$ the AdS/CFT prescription tells us precisely what to use. To construct the boundary action, we require the bulk field $\phi(x,y)$ to behave near the UV boundary as \cite{kw}:
\begin{equation}
\phi(x,y) \rightarrow  e^{(4-\Delta)ky} \varphi_0(x)+e^{\Delta k y}A(x),
\end{equation}
where the operator dimension $\Delta$ is given in (\ref{dimO}). The field $\varphi_0(x)$ is the source field, related to $\varphi^s(x)$ by an overall normalization, and $A(x)$ is interpreted as the expectation value of the CFT operator, $\langle {\cal O}(x)\rangle$, which we will not need here. This suggests the source profile is given by
\begin{equation}
g^s(y)=N_s e^{(4-\Delta) k y}
=\begin{cases}
\sqrt{\frac{2(b-1)k}{e^{2(b-1)\pi k R}-1}~}e^{bky}   \quad\quad~{\rm for} \quad b<2~,\\\\
\sqrt{\frac{2(3-b)k}{e^{2(3-b)\pi k R}-1}}~e^{(4-b)k y} \quad  {\rm for} \quad  b>2~.
\end{cases}
\label{fs}
\end{equation}
The normalization $N_s$ is chosen so that the kinetic term is canonical (as in (\ref{norm})).

The holographic meaning of the source profile can easily be understood: for large $|b|$ the source is UV localized, separated from the composite modes localized on the IR brane,  meaning the mixing is irrelevant. However, for $1<b<3$ the source profile with respect to a flat metric, $ \tilde{g}^s(y)=e^{- k y}g^s(y)$, is actually localized on the IR brane, corresponding  to relevant mixing between the source and CFT sectors. This precisely matches the mixing inferred from the operator dimension $\Delta$ (\ref{dimO}).

Note that when we are on the $(-)$ branch ($b<2$), the source is massless, so it is logical that $g^s(y)$ is identical to the zero mode profile $f^0(y)$ (\ref{0mode}). However, on the (+) branch ($b>2$), we know from holography that the source picks up a mass (\ref{M0}) and hence it must have a different profile. Let us examine the source dynamics on the ($+$) branch. Inserting the expansion (\ref{holobasis}) with the source wavefunction (\ref{fs}) into the action (\ref{a1}) and computing the overlap integral, we find
\begin{eqnarray}
S & = &\int d^5x \left[-\frac{1}{2}e^{-2ky}(g^s)^2(\partial_\mu \varphi^s)^2-\frac{1}{2}e^{-4ky}(\partial_5 g^s)^2(\varphi^s)^2\right.\nonumber \\
& &\left. \qquad\quad-\frac{1}{2}ak^2e^{-4ky}(g^s)^2(\varphi^s)^2
-bk e^{-4 k y}(g^s)^2(\varphi^s)^2\left(\delta(y)-\delta(y-\pi R)\right)+\cdots
\right],\nonumber\\
&=& \int d^4x \left[-\frac{1}{2}(\partial_\mu \varphi^s)^2-\frac{1}{2}M_s^2(\varphi^s)^2+\cdots\right]~,
\label{asource}
\end{eqnarray}
where we have defined $M_s^2$ to be
\begin{equation}
M_s^2=\frac{e^{2(2-b)\pi k R}-1}{e^{2(3-b)\pi k R}-1}4 (b-2)(b-3)k^2~.
\label{sourcemass}
\end{equation}
This matches the result from holography (\ref{M0}) except for the exponential factor. The origin of this coefficient can likely be derived from considering renormalization group running arising from source-CFT interactions. We will simply check that the correct mass eigenvalues are obtained after diagonalizing the holographic Lagrangian, which would not happen if the source had a different mass\footnote{ In particular, in the Appendix we give an analytic proof of the existence of a massless mode, and the exponential factor in (\ref{sourcemass}) is crucial in the proof.}.

One final point regarding the holographic basis deserves comment. Regardless of which basis we use, the bulk field $\phi(x,y)$ must satisfy the boundary condition (\ref{bc}). In the Kaluza-Klein basis, the profiles $f^n(y)$ satisfy this condition by definition, so clearly the bulk field $\phi(x,y)$ does as well. However, in the holographic basis, the functions $g^n(y)$ obey Dirichlet conditions on the UV boundary, and $g^s(y)$ does not satisfy the boundary condition (\ref{bc}) on the $(+)$ branch. It is possible to show, using the 4D equations of motion for the the source $\varphi^s$ and CFT composites $\varphi^n$, that the bulk field does indeed satisfy (\ref{bc}) in a nontrivial manner.

Armed with a complete definition of the holographic basis (\ref{holobasis}), we can now decompose the bulk action (\ref{a1}) and examine the elementary/composite mixing in the holographic theory.

\subsection{The eigenvalue problem}

Expanding the field in the holographic basis (\ref{holobasis}) will, by construction, produce mixing between the source $\varphi^s$ and the CFT fields $\varphi^n$. In this section we will outline the procedure for diagonalizing the system. In the end, we must reproduce the mass eigenstates derived from the Kaluza-Klein procedure (\ref{spectrum1}).

Inserting the expansion (\ref{holobasis}) into the action (\ref{a1}), we have
\begin{equation}
S=S(\varphi^s)+S(\varphi^n)+S_{mix}~,
\label{Lmix}
\end{equation}
where
\begin{eqnarray}
S(\varphi^s)&=&\int d^4x \left[-\frac{1}{2}(\partial_\mu \varphi^s)^2-\frac{1}{2}M_s^2(\varphi^s)^2\right], \\
S(\varphi^n)&=&\int d^4x \sum_{n=1}^\infty\left[-\frac{1}{2}(\partial_\mu \varphi^n)^2-\frac{1}{2}M_n^2(\varphi^n)^2\right],\\
S_{mix}&=&\int d^4x \sum_{n=1}^{\infty}\left[-z_n\partial_\mu \varphi^s \partial^\mu \varphi^n-\mu_n^2\varphi^s \varphi^n \right].\label{Lmix2}
\end{eqnarray}
The diagonal CFT masses $M_n^2$ are determined from (\ref{spectrum2}).
We see there is kinetic mixing $z_n$ and mass mixing $\mu_n^2$, both of which can be computed from wavefunction overlap integrals:
\begin{eqnarray}
z_n&=&\int_0^{\pi R}  dy~e^{-2 k y}g^s g^n, \label{kineticmix}\\
\mu_n^2 &=&\int_0^{\pi R} dy~e^{-4 k y}\left[ \partial_5 g^s \partial_5 g^n + ak^2 g^sg^n
+2 b k  g^s g^n \left(\delta(y)-\delta(y-\pi R)\right)\right]~.\label{massmix}
\end{eqnarray}
The kinetic mixing $z_n \neq 0$, which means that the functions $g^s(y)$ and $g^n(y)$ form a nonorthogonal basis.

We can represent the system more compactly in matrix notation:
\begin{equation}
{\cal L}=\frac{1}{2}\vec{ \varphi}^{\rm T} {\bf Z} \Box \vec{\varphi}-\frac{1}{2}\vec{\varphi}^{\rm T}
{\bf M}^2\vec{\varphi}~,
\end{equation}
where $\vec{\varphi}^{\rm T}=(\varphi^s, \varphi^1, \varphi^2, \cdots)$ and the mixing matrices are defined as
\begin{equation}
{\bf Z}= \left( \begin{array}{ccccc} 1 & z_1 & z_2 & z_3 &\cdots \\ z_1 & 1 & 0 & 0 & \cdots\\
z_2 & 0 & 1 & 0 & \cdots \\ z_3 & 0 & 0 & 1 & \cdots\\ \vdots & \vdots & \vdots & \vdots & \ddots \end{array} \right),\label{kmix}
\end{equation}
\begin{equation}
{\bf M}^2=  \left( \begin{array}{ccccc} M_s^2 & \mu_1^2 & \mu_2^2 & \mu_3^2 &\cdots \\ \mu_1^2 & M_1^2 & 0 & 0 & \cdots\\
\mu_2^2 & 0 & M_2^2 & 0 & \cdots \\ \mu_3^2 & 0 & 0 & M_3^2 & \cdots \\ \vdots & \vdots & \vdots & \vdots & \ddots \end{array} \right).\label{mmix}
\end{equation}

To diagonalize this system, we proceed in three steps, analogous to the $2\times 2$ problem discussed in Section 2. First we perform an orthogonal rotation in field space, $\vec{\varphi}\rightarrow {\rm \bf U} \vec{\varphi}$, which diagonalizes the kinetic portion of the Lagrangian. Second, although the resulting kinetic action is diagonal, we must additionally canonically normalize the action. We do this via a nonorthogonal diagonal matrix ${\rm \bf T} = {\rm diag}(1/\sqrt{{\rm eigenvalue}({\bf Z})})$. Altogether, we have
\begin{eqnarray}
{\bf Z} &\rightarrow &{\bf T}~{\bf U}~{\bf Z}~{\bf U}^{\rm T}~{\bf T}={\bf 1}~, \\
{\bf M}^2 &\rightarrow &{\bf T}~{\bf U}~{\bf M}^2~{\bf U}^{\rm T}~{\bf T}={\bf M'}^2~. \label{mp}
\end{eqnarray}
Third, the transformations that diagonalize the kinetic terms will create a more complicated mass matrix ${\bf M'}^2$ than initially appears in (\ref{mmix}). We must therefore perform another orthogonal field rotation, $\vec{\varphi}\rightarrow {\rm \bf V} {\rm \bf T}^{-1} {\rm \bf U} \vec{\varphi}$, which diagonalizes the mass Lagrangian,
\begin{equation}
{\bf M}^2 \rightarrow  {\bf V}~{\bf T}~{\bf U}~{\bf M}^2~{\bf U}^{\rm T}~{\bf T}~{\bf V}^{\rm T}={\bf m}^2~.
\end{equation}
If our hypothesis regarding the holographic basis is correct, the diagonalized system must match the Kaluza-Klein mass eigenbasis:
\begin{equation}
{\bf m}^2= \left( \begin{array}{ccccc} 0 & 0 & 0 & 0 &\cdots \\0  & m_1^2 & 0 & 0 & \cdots\\
0 & 0 & m_2^2 & 0 & \cdots \\ 0 & 0 & 0 & m_3^2 & \cdots\\ \vdots & \vdots & \vdots & \vdots & \ddots \end{array} \right).
\label{massbasis}
\end{equation}
We will verify that this is indeed the case in Section 5.

Finally, we can write the mass eigenstates in terms of the source and CFT fields to see precisely how much each mass eigenstate is elementary and composite. Defining $\vec{\phi}^{\rm T}=(\phi^0, \phi^1, \phi^2, \cdots)$, we have
\begin{equation}
\vec{\phi}={\bf V}~{\bf T}^{-1}~{\bf U}~\vec{\varphi}~.
\label{tran}
\end{equation}

Notice that the transformation ${\bf T}$ is not orthogonal, but rather simply a scaling of the fields. Thus, the mass eigenstates cannot be written as an orthogonal combination of source and CFT fields. It is still possible to characterize the source/CFT content for any given mass eigenstate by examining the corresponding eigenvector.

The first nontrivial check of our formalism is the existence of a zero mode, which is true if $\det {\bf M'}^2=0$. It is straightforward to compute this determinant:
\begin{eqnarray}
\det {\bf M'}^2 \propto M_s^2-\sum_{n=1}^\infty \frac{\mu_n^4}{M_n^2}~.
\label{C0}
\end{eqnarray}
We will see that $\det {\bf M'}^2=0$  is trivially satisfied on the $(-)$ branch since the source is massless and there is in fact no mass mixing. On the $(+)$ there is mass mixing as well as a massive source field, so it is certainly nontrivial that $\det {\bf M'}^2$ vanishes, as we will discuss shortly.

We are not able to offer an analytic solution to the eigenvalue problem, and leave it as an interesting open problem. Instead of an analytic diagonalization, in Section 5 we present numerical examples that our formalism is correct. In order to facilitate these calculations, we present next the analytic expressions for the kinetic and mass mixing coefficients $z_n$ and $\mu_n^2$ for each branch.

\subsubsection{$(-)$ branch}
The source is massless on the $(-)$ branch, $M_s^2=0$.  Moreover, there is no mass mixing on the $(-)$ branch. We can see this by integrating by parts in (\ref{massmix}) and using the equation of motion for $g^s(y)$, which is the same as the zero mode $f^0(y)$ and is given in (\ref{eom1}). Taking note of the boundary conditions (\ref{holobcUV}) and (\ref{holobc}), $\mu_n^2$ is easily seen to vanish. Clearly, $\det {\bf M'}^2=0$ (\ref{C0}), and there is a massless eigenstate.

Thus on the $(-)$ branch, there is only kinetic mixing. This mixing can be computed analytically. Inserting the wavefunctions $g^s$ and $g^n$ into (\ref{kineticmix}), we have
\begin{eqnarray}
z_n&=&N_sN^{{\rm CFT}}_n\int_0^{\pi R} dy~e^{b k y}\left[J_\alpha\left(\frac{M_n e^{ky}}{k}\right)+\kappa(M_n)Y_\alpha\left(\frac{M_n e^{ky}}{k}\right)\right]~,\nonumber\\
&=& \frac{N_s N^{{\rm CFT}}_n}{k}\left(\frac{M_n}{k}\right)^{\alpha-2}\int_{u(0)}^{u(\pi R)} du~u^{1-\alpha} \Big[J_\alpha(u)+\kappa(M_n)Y_\alpha(u)\Big]~, \nonumber\\
&=& -\frac{2kN_sN^{{\rm CFT}}_n}{\pi M_n^2 Y_\alpha\left(\frac{M_n}{k}\right)}~.
\label{kn-}
\end{eqnarray}
In the second line above, we have changed variables to $u=M_n e^{ky}/k$.

In fact, for most values of $b$ on the minus branch, the mixing problem is well described at low energies by the truncated problem described in Section 2. We can identify $\sin\theta=z_1$, and for $n=1,2,\dots$ use the approximate mass formula,
\begin{equation}
M_n\simeq \left(n-\frac{b}{2}+\frac{1}{4}\right)\pi k e^{-\pi k R}~,
\end{equation}
to write
\begin{equation}
\sin\theta\simeq c(b) \sqrt{\frac{b-1}{1-e^{2(1-b)\pi k R}}}~,
\end{equation}
where $c(b)$ is an $O(1)$ coefficient independent of $k$ and $R$. This simplification works well for $b<1$, where the source/CFT mixing is small, and the first CFT mass is above the IR scale. However, in the region $b\sim 2$, the CFT produces a very light composite state  and the above expression is not valid.

\subsubsection{$(+)$ branch}
On the $(+)$ branch, there is nontrivial mass mixing as well as kinetic mixing between the source and CFT sectors, and thus the eigenvalue problem is somewhat more complicated, but the diagonalization procedure is the same. The mixing coefficients $z_n$ and $\mu_n^2$ can still be computed analytically. Let us first examine the mass mixing $\mu_n^2$. Integrating by parts in (\ref{massmix}) and using the source wavefunction $g^s(y)$ (\ref{fs})  the coefficient $\mu_n^2$ can be written as a boundary term:
\begin{equation}
\mu_n^2=2(2-b)k e^{-4\pi k R}g^s(\pi R)g^n(\pi R)~.
\label{mu3}
\end{equation}
Inserting the wavefunctions (\ref{fs}) and (\ref{gcft}), we have
\begin{equation}
\mu_n^2=\frac{2 k N_sN^{{\rm CFT}}_n}{\pi Y_{\alpha+1}\left(\frac{M_ne^{\pi k R}}{k}\right)}\frac{2\alpha k}{M_n} e^{-(\alpha+1)\pi k R}~.
\label{+massmix}
\end{equation}

The kinetic mixing $z_n$ (\ref{kineticmix}) can be evaluated in a similar manner to (\ref{kn-}), and is given by
\begin{equation}
z_n=\frac{2 k N_sN^{{\rm CFT}}_n}{\pi M_n^2}\left[\frac{2\alpha k e^{-(\alpha+1)\pi k R}}{M_n Y_{\alpha+1}\left(\frac{M_ne^{\pi k R}}{k}\right)}-\frac{1}{Y_{\alpha}\left(\frac{M_n}{k}\right)}\right]~.
\label{kn+}
\end{equation}

On the $(+)$ branch, it is not readily apparent that there is a massless eigenstate since $\det {\bf M'}^2=0$ (\ref{C0}) is nontrivial. However, we show analytically in the Appendix that
\begin{equation}
\sum_{n=1}^\infty \frac{\mu_n^4}{M_n^2}=M_s^2~,
\label{musum}
\end{equation}
implying that there is indeed a massless eigenstate on the ($+$) branch.

A simple $2\times 2$ truncation will not work on the $(+)$ branch because the source field is massive (\ref{sourcemass}). To ensure an accurate diagonalization, we must include a certain minimum number of composites in the truncation such that the heaviest mass does not belong to the source field.

\subsubsection{Source/resonance mixing}
As a nontrivial check that the holographic basis correctly describes the proper interactions between the source and 
CFT sectors, consider the matrix element:
\begin{equation}
\langle \varphi^s | \frac{1}{\sqrt{Z_0}} \varphi^s {\cal O} | n \rangle
=\frac{1}{\sqrt{Z_0}} \langle \varphi^s |\varphi^s|0\rangle \langle 0 |{\cal O} |n\rangle  = \frac{F_n}{\sqrt{Z_0}}~,
\label{decay}
\end{equation}
where $Z_0=1/N_s^2$ can be derived from the correlator \cite{review}. Using the analytic expressions for $z_n$ (\ref{kn-},~\ref{kn+}) and $\mu_n^2$ (\ref{+massmix}), we can write the amplitude $F_n$ (\ref{fn1}) in the following simple form
\begin{equation}
F_n=  \sqrt{Z_0}(z_n p^2 +\mu_n^2)\bigg\vert_{p^2=-M_n^2}.
\end{equation}
The result (\ref{decay}) clearly matches that derived from (\ref{Lmix2}), since in momentum space the Lagrangian is
\begin{equation}
{\cal L}=-(z_n p^2+\mu_n^2) \varphi^s(-p)\varphi^n(p)+\dots, \nonumber \\
\end{equation}
and hence the amplitudes are identical.

\subsection{Eigenvectors}
Once we trust that the holographic basis correctly describes the elementary/composite mixing of the dual theory, we can obtain the eigenvectors directly by equating the Kaluza-Klein (\ref{kk}) and holographic (\ref{holobasis}) expansions of the bulk field $\phi(x,y)$:
\begin{equation}
\sum_{n=0}^\infty\phi^n(x)f^n(y)=\varphi^s(x)g^s(y)+\sum_{n=1}^\infty\varphi^n(x)g^n(y)~.
\end{equation}
Using the orthonormal condition (\ref{norm}), we can write the mass eigenstate in terms of the source and CFT fields:
\begin{equation}
\phi^n(x)=v^{ns}\varphi^s(x)+\sum_{n=1}^\infty v^{nm}\varphi^m(x)~,
\end{equation}
where
\begin{eqnarray}
v^{ns} & = &\int dy\, e^{-2ky} f^n(y) g^s(y)~, \label{v1}\\
v^{nm} & = &\int dy\, e^{-2ky} f^n(y) g^m(y)~. \label{v2}
\end{eqnarray}

In particular, for the massless mode $\phi^0(x)$, the integrals can be performed analytically. Consider first the ($-$) branch, $b<2$. Since $g^s(y)=f^0(y)$, the eigenvector takes a very simple form with $v^{0s}=1$, $v^{0m}=z_m$, where $z_m$ is given in (\ref{kn-}).

On the ($+$) branch, the source wavefunction (\ref{fs}) is different from the $f^0(y)$, but it is still straightforward to compute the zero mode eigenvector. Consider $v^{0s}$:
\begin{eqnarray}
v^{0s}&=&\sqrt{\frac{(3-b)}{e^{2(3-b)\pi k R}-1}}\sqrt{\frac{(b-1)}{e^{2(b-1)\pi k R}-1}}(e^{2 \pi k R}-1)~,\\
&\simeq&\begin{cases}
\sqrt{(3-b)(b-1)} \quad\quad\quad\quad\quad~{\rm for} \quad 2<b<3~,\\\\
\sqrt{(b-3)(b-1)}e^{-(b-3)\pi k R}   \quad  {\rm for} \quad  b>3~.
\end{cases}
\label{v0+}
\end{eqnarray}
This matches our expectation from the dependence of the dimension of the CFT operator ${\cal O}$ on $b$. For $2<b<3$ there is a relevant coupling between the source and CFT sectors, reflected by the fact that the source yields an order one contribution to the massless mode in (\ref{v0+}). On the other hand, the source contribution to the zero mode content is exponentially suppressed for $b>3$, consistent with our knowledge that the source/CFT interaction is irrelevant for large values of $b$.

We can also compute $v^{0n}$ for $b>2$, which is found to be
\begin{equation}
v^{0n}=\frac{-2k N_s N^{CFT}_n}{\pi M_n^2 Y_\alpha(\frac{M_n}{k})}=z_n-\frac{\mu^2_n}{M_n^2}~.
\end{equation}

For the first composite state, which has an exponentially light mass (\ref{Mapprox}), we can show for $b>3$ that $v^{01}\sim 1$. On the other hand, for the higher composite modes $n>1$, $v^{0n}$ is exponentially suppressed. Along with (\ref{v0+}), this tells us that on the (+) branch for $b>3$, the zero mode is effectively the first CFT bound state:
\begin{equation}
\phi^0(x) \sim \varphi^1(x)~.
\end{equation}
As another check, the CFT wavefunction $g^1(y)$ takes the following form for large values of $\alpha$:
\begin{equation}
g^1(y)\simeq 2 \sqrt{2(\alpha+1)k}e^{-(\alpha+1)\pi k R}e^{2ky}\sinh{\alpha k y}~.
\end{equation}
For $y>0$ this matches the zero mode wavefunction (\ref{0mode}). Note that we must expect some deviation from the zero mode profile $f^0(y)$ near $y=0$ since $g^1(y)$  obeys Dirichlet conditions at the UV boundary.

Finally, consider massive eigenmodes. On the $(-)$ branch, these modes are purely composite and contain no source field. Explicitly, since $g^s(y)=f^0(y)$, $v^{ns}=0$ by (\ref{norm}). However, the massive eigenmodes do become partly elementary on the $(+)$ branch, $v^{ns}\neq 0$.

\subsubsection{Physical interpretation of the mass eigenstates}

To define the ``compositeness'' of a mass eigenstate when the mixing involves an infinite set of composite resonances, let us focus on the massless mode, with eigenvector
\begin{equation}
\phi^0(x)=v^{0s}\varphi^s(x)+\sum_{n=1}^\infty v^{0n}\varphi^n(x)~.
\end{equation}
Mathematically, it would be natural to define the ``compositeness'' of the zero mode by the following fraction $\epsilon$ :
\begin{equation}
\epsilon=\frac{\displaystyle \sum_{n=1}^{\infty}(v^{0n})^2}{\displaystyle(v^{0s})^2+ \sum_{n=1}^{\infty} (v^{0n})^2}.
\label{comp1}
\end{equation}
We show in the Appendix that
$\sum_{n=1}^{\infty}(v^{0n})^2=1~,$
for all values of $b$. This leads to an apparent paradox. On the $(-)$ branch, we previously showed that $v^{0s}=1$ for all values of $b$. By our above definition of compositeness (\ref{comp1}), the zero mode is 50\% elementary - 50\% composite, regardless of the value of $b$. For example, the gauge boson and the graviton are equally composite by the definition $\epsilon$ above (\ref{comp1}).

To an experimenter, however, the ``compositeness'' of a particle is an energy dependent statement. As an illustration, consider a measurement of the electromagnetic form factor of the pion $F_\pi(p^2)$. This form factor can be interpreted as an effect arising from the hadronic structure of the photon \cite{bsyp}. At low energies the probe is pointlike and composed of the QED photon, while at energy scales of order 1 GeV the $\rho$ meson mixes with the elementary photon. The form factor gives us information about the structure of the ``composite'' photon, in particular the couplings between the QED photon and the QCD $\rho$ meson, which are analogous to $z_n$ and $\mu_n^2$ in our case. Of course, as we continue to increase the energy of the probe, heavier resonances can mix with the QED photon, but at low energies, $F_\pi$ will be insensitive to these resonances.

Taking this physical point of view, we would simply integrate out heavy resonances which decouple from a given physical process with a given energy scale. For instance, if our probe has energies of order the mass of the first composite state $M_1$, we would define the ``compositeness'' $\epsilon$ to be
\begin{equation}
\epsilon=\frac{(v^{01})^2}{(v^{0s})^2+  (v^{01})^2}.
\label{comp2}
\end{equation}
Indeed, if experimenters begin to probe the Standard Model partial compositeness at the LHC, (\ref{comp2}) would be the quantity they probe, not (\ref{comp1}). This resolves the previous paradox: when we say the gauge boson is more ``composite'' than the graviton, we mean that it may be possible to probe the composite structure of the gauge boson with TeV scale probes while the graviton will appear to be pointlike up to much higher scales.

\section{Numerical examples}

In this section we will provide convincing numerical evidence that the holographic basis (\ref{holobasis}) correctly describes the pure source and CFT bound state fields and the mixing between the two sectors. Although we cannot provide an analytic solution to the general eigenvalue problem, we can, to a high degree of accuracy, solve the problem numerically by considering the truncated theory, with $N$ total states $\varphi^s, \varphi^1, \dots , \varphi^{N-1}$.

The scalar field theory we are studying is actually quite versatile, in the sense that for particular values of the localization parameter, the theory mimics that of other bosonic theories with localized massless modes. For example, the graviton $h_{\mu\nu}(x,y)$ can be described by the scalar theory for $b=0$: the same equations of motion and profiles (with respect to a flat metric), the same mass eigenvalue equations and masses, and the same structure of the effective 4D Lagrangian, up to tensorial structure, can be derived from either theory. In fact, two phenomenologically relevant examples from RS model building, the graviton $(b=0)$, and the gauge boson $(b=1)$,
will be studied in this section by simply lifting the results from our scalar field theory\footnote{In fact it is possible to change the elementary/composite nature of the gauge field \cite{bulkgauge} and the graviton \cite{gpp}. Our generic scalar field theory describes the elementary/composite mixing in these cases as well.}. These examples are also pedagogical due to the varying degrees of elementary/composite mixing. To generate the Planck-weak hierarchy we will use $\pi k R \sim 34.54~$  and $k \sim 10^{15} ~ {\rm TeV} \sim 0.1 M_P~$, giving Kaluza-Klein masses of order several TeV.

These phenomenological examples are described by the $(-)$ branch, $b\leq2$. We will also demonstrate the validity of the holographic basis on the (+) branch with a few numerical examples. In these cases the source field is massive (\ref{sourcemass}) and there is mass mixing (\ref{+massmix}) in addition to kinetic mixing between the elementary and composite states. In fact there are some subtleties on this branch due to the heavy source field, which we will discuss shortly.

Finally, we will also comment on the phenomenologically important scalar examples of the radion and gauge-Higgs scalar $A_5$. These fields are Nambu-Goldstone bosons, excited by gauge currents, and as such cannot be modeled by a simple bulk scalar field.

For each case, we will generically use $\tilde{f}^0(y)$ to denote the zero mode wavefunction with respect to a flat metric. We will show the mass eigenvalues $m_n$ from the exact expression (\ref{spectrum1}), as well as the CFT masses $M_n$ (\ref{spectrum2}) and source mass $M_s$ (\ref{sourcemass}). We also present the numerical results of the mass eigenvalues, computed from the diagonalization of the mixed system (\ref{Lmix}) for truncated theories with $N=4,10,100$ states. The results match the physical masses $m_n$ astonishingly well. Finally, the transformation matrix (\ref{tran}) is shown in several cases, from which we can see the precise elementary/composite content of each mass eigenstate.

\subsection{Graviton}
The zero mode graviton $h^0_{\mu\nu}(x)$ is localized exponentially on the UV brane, with profile \cite{rs2}
\begin{equation}
\tilde{f}^0(y)\sim e^{-ky}.
\end{equation}
In correlation with this UV localization, the massless graviton in the dual theory is, for all practical purposes, the source field. This can be seen in several ways. First, if we compute that the CFT spectrum derived from Dirichlet-Neumann conditions (\ref{spectrum2}), we find that it is identical the Kaluza-Klein spectrum (\ref{spectrum1}) to
more than 14 significant figures.
Second, the transformation which diagonalizes the system is extremely close to the unit matrix: ${\bf V}~{\bf T}^{-1}~{\bf U}\sim {\bf 1}$. In particular, the massless eigenstate is written (suppressing Lorentz indices)
\begin{equation}
h^0(x)\sim h^s(x)+\sin\theta_g h^1_{CFT}(x)+\cdots,
\end{equation}
where
$\sin\theta_g\sim\theta_g \sim 2.48~e^{-\pi k R}\sim 10^{-15}$. In the case of the graviton the holographic basis is effectively equivalent to the mass eigenbasis, as we expect when the zero mode is heavily UV localized. Furthermore, while the zero mode is primarily elementary the Kaluza-Klein modes are purely composite. In particular the first Kaluza-Klein mode decomposes as
\begin{equation}
h^1(x)\sim \cos\theta_g h^1_{CFT}(x)+\cdots,
\end{equation}
where $\cos\theta_g \sim 1-\theta_g^2$. The higher Kaluza-Klein modes can similarly be written in terms
of the CFT states.

\subsection{Gauge field}

Because the graviton is localized on the UV brane, the mixing is negligible and the diagonalization is trivial. A nontrivial example is provided by the bulk gauge field. The zero mode $A^0_\mu(x,y)$ has a flat profile \cite{gaugeb1,gaugeb2}:
\begin{equation}
\tilde{f}^0(y)=\frac{1}{\sqrt{\pi R}}.
\end{equation}
The field is not localized in the extra dimension and thus we expect it to have a significant composite mixture in the dual theory. The dual interpretation of bulk gauge fields is discussed in \cite{pheno1,pheno4}. Table~\ref{table2} shows that the diagonal CFT masses differ from the eigenmasses to a much greater extent than in the graviton case.
\begin{table}[ht]
\centerline{
\begin{tabular}{|c||c|c||c|c|c|}
	\hline
$n$ & $m_n$    &   $M_n$  & $N=4$ & $N=10$ &  $N=100$   \\
	\hline \hline
0   & 0 & 0 & 0 & 0  & 0\\
1   &  2.44751138 & 2.40188495  & 2.44755043 &  2.44751619 & 2.44751142 \\
2   & 5.56073721 & 5.51332800 & 5.56098310 & 5.56076402 & 5.56073743 \\
3   & 8.69131408 & 8.69131465 & 8.69219991 & 8.69138348 & 8.69131465  \\
	\hline
\end{tabular}}
\caption{Gauge field, $A_\mu$ $(b=1)$: a comparison of physical masses $m_n$ with CFT masses $M_n$ and mass eigenvalues numerically computed from the $N=4, 10, 100$ truncated eigenvalue problem. }
\label{table2}
\end{table}
Also shown are the eigenvalues computed from the truncated $N \times N$ problem. Notice that the more states we include (larger $N$), the more precisely the eigenvalues match those derived directly from the Kaluza-Klein spectrum (\ref{spectrum1}). This is a good indicator that our procedure is working and the holographic basis indeed corresponds to the pure elementary and pure composite fields in the dual theory.

The transformation matrix which diagonalizes the gauge field action is
\begin{equation}
\left( \begin{array}{c} A_\mu^0  \\ A_\mu^1 \\ A_\mu^2 \\ \vdots \end{array} \right)
= \left( \begin{array}{rrrr} 1 & -0.19 & 0.13 &\cdots \\
 0 & -0.98 & -0.03 & \cdots \\
  0 & 0.01 & -0.99 &\cdots \\
\vdots & \vdots & \vdots & \ddots
\end{array} \right)
 \left( \begin{array}{c} A_\mu^s \\ A_\mu^{1(CFT)} \\ A_\mu^{1(CFT)} \\ \vdots \end{array} \right).
\label{tranamu}
\end{equation}
The zero mode gauge field is primarily an elementary field.
The massive eigenstates, on the other hand, are comprised of purely composite fields, with no elementary mixture. In fact an approximate analytic expression can again be written for the $N=2$
case, leading to
\begin{eqnarray}
   A_\mu^0  &\sim& A_\mu^s +\sin\theta_A A_\mu^{1(CFT)}+\dots~,\nonumber\\
      A_\mu^1  &\sim& \cos\theta_A A_\mu^{1(CFT)}+\dots~,
 \end{eqnarray}
 where $\sin\theta_A \sim -1.13/\sqrt{\pi k R}$.

Because the zero mode is flat, we might have expected a massless eigenstate containing roughly half source and half composite content. However, localization of the zero mode provides only a rough guide to the holographic theory. In fact, the mixing between the elementary gauge field and the corresponding CFT current $J_\mu^{CFT}$ is marginal since $\Delta_J=3$ (\ref{dimO}), explaining why the zero mode is primarily elementary.

\subsection{Scalar fields}
\subsubsection{$b=2$}

An interesting instructive example occurs for a scalar field with $b=2$.
The zero mode is localized on the IR brane,
\begin{equation}
\tilde{f}^0(y)\sim e^{ky}.
\label{b2profile}
\end{equation}
At the point $b=2$, the dimension of the operator ${\cal O}$ takes its lowest value, $\Delta=2$, corresponding to strong mixing between the two sectors. We should therefore expect that the zero mode is half elementary and half composite.  Using the holographic basis, we can see that this is the case.

First, it is clear that the diagonalization procedure works from Table~\ref{table3}. The computed eigenvalues from the $N\times N$ problem become successively closer to the true eigenvalue $m_n$ as we increase $N$. An interesting feature in this case is that the eigenmasses and the diagonal CFT masses differ greatly level by level, in contrast to the graviton and gauge boson cases. It is clear what is happening: the CFT is producing a very light state, indicating that the massless mode will contain a large CFT component. Furthermore, comparing the diagonal masses, it is apparent that the $n$th Kaluza-Klein mode will be composed primarily of the $(n+1)$th CFT state.
\begin{table}[ht]
\centerline{
\begin{tabular}{|c||c|c||c|c|c|}
	\hline
$n$ & $m_n$    &   $M_n$  & $N=4$ & $N=10$ &  $N=100$   \\
	\hline \hline
0   & 0             & 0               & 0 & 0  & 0\\
1   & 3.82701899632 &  0.24297921275  & 3.82703429186 &  3.82701915239 & 3.82701899633  \\
2   & 7.00700617376 &  3.87522598167  & 7.00725928486 &  7.00700795318 & 7.00700617389  \\
3   & 10.1610259641 &  7.05530379320  & 69.6177731631 &  10.1610342131 & 10.1610259647  \\
	\hline
\end{tabular}}
\caption{Scalar field $(b=2)$: A comparison of physical masses $m_n$ with CFT masses
$M_n$ and mass eigenvalues numerically computed from the $N=4, 10, 100$ truncated eigenvalue problem. }
\label{table3}
\end{table}

Now let us examine the field content of the mass eigenstates. In terms of the source and CFT fields, the mass eigenstates are
\begin{equation}
\left( \begin{array}{c} \phi^0  \\ \phi^1 \\ \phi^2 \\ \phi^3\\ \vdots \end{array} \right)
= \left( \begin{array}{rrrrr}
 1 &  -0.99994 & 0.00996  &-0.00409  & \cdots \\
 0 &  -0.01005 &-0.99980  & 0.01440   & \cdots \\
 0 &  -0.00401 & 0.01462  & 0.99972   & \cdots \\
 0 &  -0.00229 & 0.00679  &-0.01526    & \cdots \\
\vdots & \vdots & \vdots  &\vdots & \ddots
\end{array} \right)
 \left( \begin{array}{c} \varphi^s \\ \varphi^{1} \\ \varphi^{2} \\ \varphi^{3} \\ \vdots \end{array} \right).
\label{trana5}
\end{equation}
For the massless eigenstate, we see that the first excited CFT field contributes almost as much as the source field. In fact $b=2$ is the transition point for which the source field provides the single most dominant contribution to the zero mode. The fact that $b=2$ is a special point was also noticed in the Schrodinger potentials~\cite{schrod}.

\subsubsection{Composite scalars}
Another phenomenological example for bosons is provided by the Higgs boson in Randall-Sundrum models \cite{rs}, which lives on the IR brane. The Higgs boson is purely a composite of the CFT, with no elementary component, which can be seen by taking the limit $b\rightarrow \infty$ in (\ref{v0+}). However, to check that our formalism works on the (+) branch, which is qualitatively different than the $(-)$ branch due to a massive source (\ref{sourcemass}) as well as mass mixing (\ref{+massmix}), we will study two examples for $b>2$. Numerically, it is much more difficult to analyze the (+) branch because the source field becomes massive. Eq. (\ref{sourcemass}) shows that almost immediately for $b>2$, the source picks up a mass of order $k$. To properly analyze the mixing and compute the correct mass eigenvalues, we must include at least as many composites so that the heaviest state in the $N\times N$ problem is not the source. For the large warping necessary to solve the hierarchy problem, this quickly becomes impractical since CFT masses are of order TeV rather than $k$. Luckily, we can compute the eigenvectors analytically as shown in Section 4.

To demonstrate that the holographic basis is correct on the (+) branch, we will first study the case $b=2.05$, using phenomenological values for $k$ and $R$. In this case, the mass of the source is comparable to the lowest lying CFT states, $M_s\sim 2.5$ TeV, and thus can easily be analyzed for relatively small $N$. Table~\ref{table4} shows clearly that the holographic basis is working.

\begin{table}[ht]
\centerline{
\begin{tabular}{|c||c|c||c|c|c|}
	\hline
$n$ & $m_n$    &   $M_n$  & $N=4$ & $N=10$ &  $N=100$   \\
	\hline \hline
0   & 0           & 2.502475807 & 0.001287282 &  0.000726358 & 0.000215505  \\
1   & 4.053119994 & 0.086364624 & 4.053121203 &  4.053120426 & 4.053120032  \\
2   & 7.367700516 & 4.059345524 & 7.367705124 &  7.367700972 & 7.367700556  \\
3   & 10.65244351 & 7.374205523 & 183.1948315 &  10.65244407 & 10.65244356  \\
	\hline
\end{tabular}}
\caption{Scalar field on $(+)$ branch $(b=2.05)$: a comparison of physical masses $m_n$ with CFT masses $M_n$ and mass eigenvalues numerically computed from the $N=4, 10, 100$ truncated eigenvalue problem. }
\label{table4}
\end{table}
Notice in particular that the diagonalized zero mode mass is nonzero for finite $N$. This shouldn't surprise us for the truncated $N \times N$ problem, as we know that the determinant of the mass matrix involves an infinite sum (\ref{C0}). The analytic proof that the general problem contains a massless eigenstate is given in the Appendix. We can be confident of the validity of the holographic basis since as we increase $N$, the diagonalized eigenvalue tends toward zero. Also, the massive eigenvalues computed from the diagonalization match the physical masses ever more precisely as we increase the number of states in the truncated theory.

The first excited CFT state now provides the dominant contribution to the zero mode in the holographic basis as is seen from the transformation matrix:
\begin{equation}
\left( \begin{array}{c} \phi^0  \\ \phi^1 \\ \phi^2 \\ \varphi^3\\ \vdots \end{array} \right)
= \left( \begin{array}{rrrrr}
 0.99987 & \sim 1   & -0.0012   & 0.0005    & \cdots \\
 0.04698 & -0.00118 & -0.999997 & 0.00184   &\cdots \\
-0.01035 &  0.00049 & -0.001839 & -0.999995 &\cdots \\
 0.01018 & -0.00029 &  0.000876 & -0.00198  & \cdots \\
\vdots & \vdots & \vdots &\vdots & \ddots
\end{array} \right)
 \left( \begin{array}{c} \varphi^s \\ \varphi^1 \\ \varphi^2 \\ \varphi^3 \\ \vdots \end{array} \right).
\label{tranphi1}
\end{equation}

Next, we analyze an example where the zero mode is largely a composite state, $b=4$. The zero mode is extremely localized on the IR brane in this case, and as mentioned a moment ago, the source receives a large mass of order $k$. It is impractical to use the phenomenological values for $k$ and $R$ to analyze this case numerically since the source mass would be Planck scale, and thus an extremely large number of composite states would have to be included in the finite $N$ problem to ensure accurate results. Instead, we will assume a far more modest warping, using $\pi k R \sim 1$ and $k \sim 1 ~{\rm TeV}$,
which makes the source mass comparable to the composite masses, thus allowing for a numerical solution with only a small number of composite states in the truncation. The numerical results are presented in Table~\ref{table5}.
\begin{table}[ht]
\centerline{
\begin{tabular}{|c||c|c||c|c|c|}
	\hline
$n$ & $m_n$    &   $M_n$  & $N=4$ & $N=10$ &  $N=100$   \\
	\hline \hline
0   &  0         &  3.013767   &  0.013504  & 0.007471  & 0.002366  \\
1   &  2.454013  &  0.252420   &  2.457739  & 2.454435  & 2.454042 \\
2   &  4.055413  &  2.784510   &  4.097470  & 4.056788  & 4.055450  \\
3   &  5.765917  &  4.601716   &  7.790989  & 5.769759  & 5.765952 \\
	\hline
\end{tabular}}
\caption{Scalar field on $(+)$ branch $(b=4)$ with weak warping: a comparison of physical masses $m_n$ with CFT masses $M_n$ and mass eigenvalues numerically computed from the $N=4, 10, 100$ truncated eigenvalue problem. }
\label{table5}
\end{table}
Again, as we increase the number of states $N$, we see that the diagonalized eigenvalues approach the physical masses. In terms of the source and composite fields, the mass eigenstates are
\begin{equation}
\left( \begin{array}{c} \phi^0  \\ \phi^1 \\ \phi^2 \\ \varphi^3\\ \vdots \end{array} \right)
= \left( \begin{array}{rrrrr}
  0.5932 &  0.9979    & -0.0418   & -0.0273  &  \cdots \\
 -0.7064 &  0.0544    &  0.9536    & 0.1944  &  \cdots \\
  0.2964 & -0.0281    &  0.2698   & -0.8852  &  \cdots \\
  0.1952 & -0.0154   &   0.1021   &  0.3848  &  \cdots \\
\vdots & \vdots & \vdots &\vdots & \ddots
\end{array} \right)
 \left( \begin{array}{c} \varphi^s \\ \varphi^1 \\ \varphi^2 \\ \varphi^3 \\ \vdots \end{array} \right).
\label{tranphi2}
\end{equation}
As we expect the zero mode is now primarily a composite field, with a much smaller admixture of source field. The massive eigenstates now contain a percentage of elementary content, different from the $(-)$ branch.

\subsection{Nambu-Goldstone modes of broken symmetry}
The final bosonic example from realistic warped models is provided by the scalar radion and $A_5$ fields. These massless fields have the following profile:
\begin{equation}
\tilde{f}^0(y)\sim e^{ky},
\end{equation}
which is identical to the zero mode of the scalar $b=2$ case (\ref{b2profile}). Thus, we might naively believe that our previous analysis of the mixing for the case $b=2$ can be applied to the radion and $A_5$. This is incorrect, and in fact these gauge degrees of freedom cannot be modeled by a simple bulk scalar field. The correct interpretation of these zero modes is that they are Nambu-Goldstone bosons of a broken symmetry, and are purely composite bound states of the CFT.

The holographic interpretation of the radion is presented in \cite{pheno1,pheno2}. The presence of an IR brane and the dual spontaneous conformal symmetry breakdown at low energies is accompanied by the appearance of the radion, the Nambu-Goldstone boson of conformal symmetry breaking.  In particular \cite{pheno2} found a massless pole in the transverse-traceless energy-momentum two-point function below the scale of conformal symmetry breaking, which can only be ascribed to the radion mode.

Similarly, the zero mode $A_5$ is a Nambu-Goldstone boson associated with the breaking of a global CFT symmetry \cite{hpgb}, with global symmetry current $J^{CFT}_\mu$. Let us examine the two point function $\langle J_\mu J_\nu\rangle$, and see that a pole at $p^2=0$ exists as well for this case. The calculation parallels that of \cite{pheno1}, except that for our case, we must apply Dirichlet conditions at the IR brane in order to obtain an $A_5$ zero mode. It is precisely this Dirichlet condition that corresponds to breaking the global symmetry with strong CFT dynamics. The approximate expression for the correlator at low energies $p \ll ke^{-\pi k R}$ is
\begin{eqnarray}
\langle J_\mu J_\nu\rangle(p) & \simeq & (p^2\eta_{\mu\nu}-p_\mu p_\nu) \frac{1}{g_5^2 k} \Big[\log{(i p/2k)}+\gamma-
\frac{\pi Y_1\left(i p e^{\pi k R}/k\right)}{2 J_1\left( i p e^{\pi k R}/k\right)}\Big] ,\\
&\simeq &  (p^2\eta_{\mu\nu}-p_\mu p_\nu)\frac{2(k e^{- \pi k R})^2}{g_5^2 k}\frac{1}{p^2}+\dots,
\end{eqnarray}
and we see the massless pole corresponding to the exchange of the Nambu-Goldstone mode $A_5$. Contrast this with the calculation of \cite{pheno1}, where Neumann boundary conditions are applied at the IR brane and no such pole appears. This is to be expected since a Neumann condition is dual to an unbroken global CFT symmetry, and a Nambu-Goldstone boson should not appear in the spectrum.

Again, we emphasize that the holographic interpretation of these modes is that they are 100\% composite. The radion and $A_5$ do not mix with an external source field, so clearly no holographic basis is necessary.

\section{Conclusion}

Theories in a slice of AdS$_5$ have a dual description in terms of bound states of a strongly coupled CFT mixing with an elementary dynamical source field. We have shown how to quantitatively describe this mixing using the holographic basis. Rather than expand the bulk field directly in mass eigenstates, we expand the field in terms of a pure source field and pure CFT composite fields. The effective Lagrangian contains in general both kinetic and mass mixing. We have shown how to diagonalize this Lagrangian, and provided numerical examples that it leads correctly back to the mass eigenbasis. We have also demonstrated analytically that there is a massless eigenstate. Although we focused on bosonic fields in this paper, the holographic basis can be used to describe fermions as well~\cite{inprep}.

The holographic basis provides an important entry in the AdS/CFT ``dictionary'', allowing one to understand the  elementary/composite interpretation of warped models in an explicit quantitative manner.  Any bulk theory containing zero modes is dual to the existence of an elementary sector, and in these cases our formalism is very useful. 
Warped models of electroweak symmetry breaking and supersymmetry breaking 
(see Ref.~\cite{review} for a review) provide relevant examples of beyond-the-Standard Model scenarios where the holographic basis can be applied to help quantify the dual interpretation. In particular for collider physics nontrivial form factors of partially composite particles could be measured at the LHC (or eventually at the ILC), and these can be conveniently computed in the holographic basis. The holographic basis is manifest in AdS/QCD models with pure QCD states interacting with elementary states of QED and gravity.

Theoretically, our formalism can be extended
to account for brane-localized kinetic terms, which could modify the holographic wavefunctions and affect
the elementary/composite content. Furthermore, it seems unlikely that this tool is exclusive to pure AdS$_5$ geometries, and therefore it would be useful to formulate a holographic basis for more general geometries, which are asymptotically AdS in the UV region. Finally it would be interesting to identify our holographic basis in warped string compactifications where the underlying dual 4D theory can in principle be determined.

\section*{Acknowledgements}
We thank Jay Hubisz and Alex Pomarol for helpful discussions. T.G. acknowledges the hospitality of
the CERN Theory Division
where part of this work was done. This work was supported in part by a Department of Energy grant DE-FG02-94ER40823 at the University of Minnesota, and an award from Research Corporation.

\newpage

\appendix
\def\theequation{\thesection.\arabic{equation}}
\setcounter{equation}{0}

\section{Sum rules}
We will present some useful sum rules by using the completeness of the CFT eigenfunctions:
\begin{equation}
\sum_{n=1}^\infty g^n(y)g^n(y')=e^{2ky}\delta(y-y')~,
\label{complete}
\end{equation}
where the CFT eigenfunctions are defined in (\ref{gcft}).
\vspace{0.4cm}

\noindent
{\bf Eigenvector coefficient sum rule}
\vspace{0.2cm}

The coefficients $v^{0n}$ (\ref{v2}) satisfy the following sum rule:
\begin{eqnarray}
\sum_{n=1}^\infty (v^{0n})^2 & = & \sum_{n=1}^\infty \int_0^{\pi R} dy~e^{-2ky}f^0(y)g^n(y) \int_0^{\pi R} dy' e^{-2ky'}f^0(y')g^n(y')~,\nonumber\\
&=&\int_0^{\pi R} dy~e^{-2 ky} (f^0(y))^2~,\nonumber\\
&=&~1~.
\end{eqnarray}
In the second line we have used the completeness relation (\ref{complete}) to perform the integral
over $y'$ and in the third line the integral follows from (\ref{norm}).
\vspace{0.4cm}

\noindent
{\bf Existence of a massless mode}
\vspace{0.2cm}

Next we present the proof of the existence of a massless eigenstate on the (+) branch as mentioned in (\ref{musum}). This follows from proving the following relation:
\begin{equation}
\sum_{n=1}^\infty \frac{\mu_n^4}{M_n^2}=M_s^2~,
\label{musum2}
\end{equation}
which implies that the determinant of the mass matrix (\ref{C0}) vanishes and a massless eigenstate exists.

The proof of (\ref{musum2}) proceeds as follows:
We first rewrite the boundary term in (\ref{massmix}) as a full derivative
\begin{equation}
\int_0^{\pi R} dy~2bk e^{-4ky}g^s g^n (\delta(y)-\delta(y-\pi R))=\int_0^{\pi R} dy~ \partial_5(-bk e^{-4ky}g^sg^n).
\end{equation}
Using the source wavefunction (\ref{fs}) and defining $\tilde{g}^n(y)=e^{-bky}g^n(y)$,
the mass mixing can be written in the following simple form:
\begin{equation}
\mu_n^2=\int_0^{\pi R}  dy~2(2-b)k e^{(b-4)ky}g^s\partial_5 \tilde{g}^n.
\label{mu1}
\end{equation}
Notice that $\tilde{g}^n(y)$ obeys purely Neumann boundary conditions at $y=\pi R$. Furthermore, we can write the equation of motion for $\tilde{g}^n(y)$ as
\begin{equation}
\partial_5e^{2(b-2)ky}\partial_5 \tilde{g}^n=-M_n^2e^{2(b-1)ky}\tilde{g}^n.
\end{equation}
Integrating this equation, we find an expression for $\partial_5 \tilde{g}^n(y)$:
\begin{equation}
\partial_5 \tilde{g}^n(y)=M_n^2 e^{-2(b-2)ky}\int_y^{\pi R} dy' e^{2(b-1)ky'}\tilde{g}^n(y')~,
\end{equation}
which can then be substituted into (\ref{mu1}), thus removing
all partial derivatives of $g^n(y)$ in the expression for $\mu_n^2$:
\begin{equation}
\mu_n^2= \int_0^{\pi R} dy~2(2-b)k M_n^2 e^{-bky}g^s(y)\left[ \int_y^{\pi R} dy' e^{2(b-1)ky'}\tilde{g}^n(y')\right].
\label{mu2}
\end{equation}

Multiplying together (\ref{mu2}) and (\ref{mu3}), the factor $M_n^2$ cancels in the sum (\ref{musum2}) and we can use the completeness relation (\ref{complete}) to perform the $y'$ integral. We finally
obtain
\begin{eqnarray}
\sum_{n=1}^\infty \frac{\mu_n^4}{M_n^2} & = &
4(b-2)^2 k^2 e^{(b-4)\pi k R}g^s(\pi R)\int_0^{\pi R}dy~ e^{-b ky}g^s(y)~, \nonumber \\
 & = &
\frac{e^{2(2-b)\pi k R}-1}{e^{2(3-b)\pi k R}-1} 4 (b-2)(b-3)k^2 = M_s^2~.
\end{eqnarray}
Thus, the determinant (\ref{C0}) is exactly zero in the infinite dimensional case, and we have proved the existence of a massless mode.



\begin{thebibliography}{99}


\bibitem{ads1}
J.~M.~Maldacena,
Adv.\ Theor.\ Math.\ Phys.\  {\bf 2} (1998) 231
[Int.\ J.\ Theor.\ Phys.\  {\bf 38} (1999) 1113]
[arXiv:hep-th/9711200].

\bibitem{ads2}
S.~S.~Gubser, I.~R.~Klebanov and A.~M.~Polyakov,
Phys.\ Lett.\ B {\bf 428} (1998) 105
[arXiv:hep-th/9802109].

\bibitem{ads3}
E.~Witten,
Adv.\ Theor.\ Math.\ Phys.\  {\bf 2} (1998) 253
[arXiv:hep-th/9802150].

\bibitem{rs}
L.~Randall and R.~Sundrum,
Phys.\ Rev.\ Lett.\  {\bf 83}, 3370 (1999)
[arXiv:hep-ph/9905221].

\bibitem{pheno1}
N.~Arkani-Hamed, M.~Porrati and L.~Randall,
JHEP {\bf 0108} (2001) 017
[arXiv:hep-th/0012148].

\bibitem{pheno2}
R.~Rattazzi and A.~Zaffaroni,
JHEP {\bf 0104} (2001) 021
[arXiv:hep-th/0012248].

\bibitem{pheno3}
M.~Perez-Victoria,
JHEP {\bf 0105} (2001) 064
[arXiv:hep-th/0105048].

\bibitem{gp}
T.~Gherghetta and A.~Pomarol,
  Nucl.\ Phys.\  B {\bf 586}, 141 (2000)
  [arXiv:hep-ph/0003129].

\bibitem{review}
T.~Gherghetta,
[arXiv:hep-ph/0601213].

\bibitem{inprep}
B.~Batell and T.~Gherghetta, in preparation.

\bibitem{crss}
  R.~Contino, T.~Kramer, M.~Son and R.~Sundrum,
  arXiv:hep-ph/0612180.

\bibitem{zzp}
K~.~S.~Babu, C.~Kolda, J.~March-Russell,
Phys.\ Rev.\ D {\bf },
[arXiv:hep-ph/9710441].

 \bibitem{thooft}
G.~'t Hooft,
Nucl.\ Phys.\  B {\bf 72}, 461 (1974);
Nucl.\ Phys.\  B {\bf 75}, 461 (1974).

\bibitem{witten}
E.~Witten,
Nucl.\ Phys.\  B {\bf 160}, 57 (1979).

\bibitem{qcd1}
J.~Erlich, E.~Katz, D.~T.~Son, M.~A.~Stephanov,
Phys.\ Rev.\ Lett.\  {\bf 95} 261602 (2005)
[arXiv:hep-ph/0501128].

\bibitem{qcd2}
L.~Da Rold, A.~Pomarol,
Nucl. Phys. B {\bf 721} 79 (2005),
[arXiv:hep-ph/0501218].

\bibitem{gp03}
 T.~Gherghetta and A.~Pomarol,
 Phys.\ Rev.\  D {\bf 67}, 085018 (2003)
 [arXiv:hep-ph/0302001].

\bibitem{kw}
I. R. Klebanov and E. Witten,
Nucl.\ Phys.\ B {\bf 556}, 89 (1999)
[arXiv:hep-ph/9905104].

 \bibitem{bsyp}
  T.~H.~Bauer, R.~D.~Spital, D.~R.~Yennie and F.~M.~Pipkin,
  Rev.\ Mod.\ Phys.\  {\bf 50}, 261 (1978)
  [Erratum-ibid.\  {\bf 51}, 407 (1979)].

\bibitem{bulkgauge}
B.~Batell and T.~Gherghetta,
Phys.\ Rev.\ D {\bf 73}, 045016 (2006)
[arXiv:hep-ph/0512356];
Phys.\ Rev.\  D {\bf 75}, 025022 (2007)
[arXiv:hep-th/0611305].

\bibitem{gpp}
T.~Gherghetta, M.~Peloso and E.~Poppitz,
Phys.\ Rev.\ D {\bf 72}, 104003 (2005)
[arXiv:hep-th/0507245].

\bibitem{rs2}
L.~Randall and R.~Sundrum,
Phys.\ Rev.\ Lett.\  {\bf 83} 4690 (1999)
[arXiv:hep-ph/9906064].

\bibitem{gaugeb1}
H.~Davoudiasl, J.~L.~Hewett and T.~G.~Rizzo,
Phys.\ Lett.\ B {\bf 473}, 43 (2000)
[arXiv:hep-ph/9911262].

\bibitem{gaugeb2}
A.~Pomarol,
Phys.\ Lett.\ B {\bf 486}, 153 (2000)
[arXiv:hep-ph/9911294].

\bibitem{pheno4}
K.~Agashe and A.~Delgado,
Phys.\ Rev.\ D {\bf 67}, 046003 (2003)
[arXiv:hep-th/0209212].

\bibitem{schrod}
   B.~Batell and A.~Larkoski,
  arXiv:hep-th/0610218.

\bibitem{hpgb}
R.~Contino, Y.~Nomura and A.~Pomarol,
Nucl.\ Phys.\ B {\bf 671}, 148 (2003)
[arXiv:hep-ph/0306259].



\end{thebibliography}
\end{document}